\documentclass[12pt,english]{article}
\usepackage{babel}
\usepackage{amsfonts,epsfig}
\title{Effective interactions and prospects for a resolution\\ of the fundamental cosmological problems in the quantum gravity}
\author{{B.A. Arbuzov and I.V. Zaitsev}\\
{Lomonosov Moscow State University}\\ {Leninskie gory 1, 119991 Moscow, Russia}
}
\date{}
\begin{document}
\maketitle
\begin{quote}
A possible effective interaction in the quantum gravity is considered. The compensation equation for a spontaneous generation of this interaction is
shown to have a non-trivial solution.
Would be consequences of a possible existence of effective interactions in the gravity theory are discussed. An example of running gravitational coupling is
presented, which corresponds to a description of effects, which nowadays are
prescribed to a dark mater and to a dark energy.


\end{quote}
\maketitle

\section{An effective three-graviton interaction}

Due to well-known problems of the dark matter and the dark energy numerous possibilities of modified gravity are considered (see, e.g. review~\cite{review} and recent work~\cite{dolgov}). This approach assumes existence of new effective interactions of the gravitational field in addition to the fundamental Einstein-Hilbert Lagrangian.

In the present work we would discuss a possibility of anomalous gravitation interaction in terms of non-perturbative effects of the Einstein-Hilbert gravity. For the purpose we rely on the approach induced by N.N. Bogoliubov compensation principle~\cite{Bog1,Bog2}.
In works~\cite{BAA04} - \cite{AZ13}, this approach
was applied to studies of a spontaneous generation of effective non-local interactions in renormalizable gauge theories. The approach is described in full in book~\cite{book}.
In particular, papers~\cite{BAA09,AZ11}  deal with an application of the approach to the electro-weak interaction and a possibility of spontaneous generation of effective anomalous three-boson interaction of the form
\begin{eqnarray}
& &\frac{g\,\lambda}{3!\,M_W^2}\,F\,\epsilon_{abc}\,W_{\mu\nu}^a\,W_{\nu\rho}^b\,W_{\rho\mu}^c\,;
\label{FFF}\\
& &W_{\mu\nu}^a\,=\,
\partial_\mu W_\nu^a - \partial_\nu W_\mu^a\,+g\,\epsilon_{abc}W_\mu^b W_\nu^c\,.\nonumber
\end{eqnarray}
where $g \simeq 0.65$ is the electro-weak coupling. Here $F(p_i)$ is a form-factor, which guarantees effective interaction~(\ref{FFF}) acting in a limited region of the momentum space. This form-factor is uniquely defined by the compensation equation of Bogoliubov approach. The approach gives unique results for physical parameters, so we have none adjusting parameter in the scheme. Would-be existence of effective interaction~(\ref{FFF}) leads to important non-perturbative effects in the electro-weak interaction~\cite{BAA09, AZPR,AZ15}. Note, that interaction~(\ref{FFF}) was considered for long time on phenomenological grounds~\cite{Hag1,Hag2}.

We would take interaction~(\ref{FFF}) as a leading hint for choosing of an effective interaction in the gravity theory. Considering links between vector non-abelian gauge theories and the gravity theory, one easily see that
gauge field $W^a_{\mu\nu}$ plays the same role as Riemann curvature tensor
$R^m_{n\,\mu\,\nu}$. Thus the anomalous interaction which is strictly
analogous to interaction~(\ref{FFF}) is the following
\begin{eqnarray}
& &L_{eff}^{G}\,=\,\frac{G}{2!}\,F\,\sqrt{-g}\,\epsilon^{n b d c}\,R_{m n \mu \nu}\,R_{a b \lambda \rho}\,R_{c d \sigma \delta} g^{\,m a}\,g^{\,\nu \lambda}\,g^{\,\rho \sigma}\,g^{\,\mu \delta}\,;\nonumber\\
& &R^s_{n \mu \nu}=\frac{
\partial \Gamma^s_{n \nu}}{\partial x_\mu} - \frac{\partial \Gamma^s_{n \mu}}{\partial x_\nu}\,+\Gamma^s_{r \mu}\,\Gamma^r_{n \nu}\,-\,\Gamma^s_{r \nu}\,\Gamma^r_{n \mu}\,;\quad R_{m n \mu \nu}=g_{m s}\,R^s_{n \mu \nu}\,;
\label{GGG}\\
& &\Gamma^i_{k l}\,=\,\frac{1}{2}\,g^{im}\biggl(\frac{\partial g_{m k}}{\partial x^l}\,+\,\frac{\partial g_{m l}}{\partial x^k}\,-\,\frac{\partial g_{k l}}{\partial x^m}\biggr)\,.\nonumber
\end{eqnarray}
Here $F$ is again some form-factor to be defined by a compensation equation.
We introduce this equation in the first approximation in what follows.
We define the Lorentz structure of the anomalous three-graviton vertex,
being quite lengthy,  by
calculations with application of FORM. We also use the standard Feynmann rules for the quantum gravity~\cite{DW,FP}.

It is important to emphasize, that wouldbe interaction~(\ref{GGG}) violates both the spatial P-invariance and the temporal T-invariance. Thus it might
influence qualitative features of the Universe evolution. In particular,
a violation of
the T-invariance is necessary for an origin of the baryon asymmetry
of the Universe~\cite{saharov}.
 \begin{figure}
\includegraphics[scale=1.2,width=15cm]{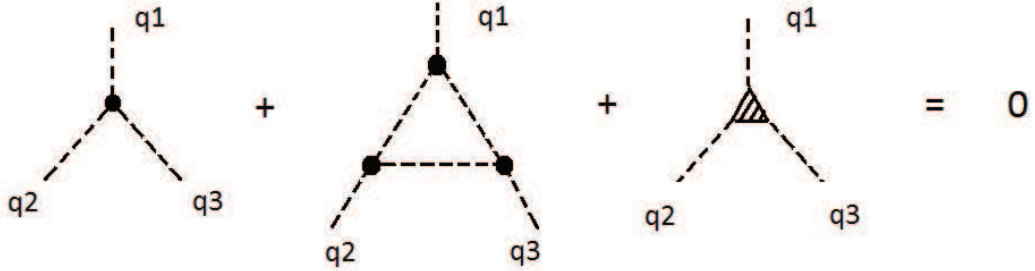}
\caption{A diagram representation of the compensation
equation in the first approximation. Dotted lines correspond to gravitons, a black spot represents interaction~(\ref{GGG}), the striped triangle represents a contribution of the Standard Model diagrams.}
\label{fig:compenG}
\end{figure}

Now let us turn to the compensation equation, which firstly answers the most important question, if interaction~(\ref{GGG}) can be spontaneously
generated, and secondly, in case of affirmative answer to the first question,
provides form-factor $F(p_i)$.

We start with the standard Einstein-Hilbert Lagrangian and expression $L_{SM}$, describing gauge interactions of the Standard Model.
\begin{equation}
L_0\,=\,\frac{1}{8\,\pi\,\kappa^2}\sqrt{-g}\,R\,+\,L_{SM}\,.\label{SM}
\end{equation}
Then we apply to expression~(\ref{GGG}) the add-subtract procedure, which in details is explained in~\cite{book}
\begin{eqnarray}
& &L = L'_0\,+\,L'_{int}\,;\quad L'_0 = L_0\,-\,L_{eff}^{G}\,; \label{eq:L'WB0}\\
& &L'_{int}\,=\,L_{eff}^{G}\,. \label{eq:L'WBInt}
\end{eqnarray}

Now let us formulate the compensation equation.
We are to demand, that considering the theory with Lagrangian
$L'_0$~(\ref{eq:L'WB0}), all contributions to three-graviton
connected vertices, corresponding to Lorentz structure~(\ref{GGG}) are summed up to zero.  That is the undesirable interaction part in the wouldbe free Lagrangian~(\ref{eq:L'WB0}) is compensated. Then we are rested with
interaction~(\ref{GGG}) only in the proper place~(\ref{eq:L'WBInt}). In a diagram form this
demand in the first approximation is presented in Figure~\ref{fig:compenG}.

The corresponding integral
equation with integrations in the Euclid momentum space is obtained also by FORM calculations and turns to be
the following

\begin{eqnarray}
& &F(x) = F_{0G} +  \frac{3 G^2}{16 \pi^2}\biggl(- \frac{1}{x^3}\int_0^x y^7 F(y) dy + \frac{4}{x^2}\int_0^x y^6 F(y) dy- \frac{3}{x}\int_0^x y^5 F(y) dy +\nonumber\\
& &
 \nonumber\\
& &8 \int_0^x y^4 F(y) dy + 18 \int_x^\infty y^4 F(y) dy -25 x \int_x^\infty y^3 F(y) dy  + \label{EQ}\\
& &24 x^2 \int_x^\infty y^2 F(y) dy-
11 x^3 \int_x^\infty y F(y) dy + 2 x^4 \int_x^\infty  F(y) dy\biggr)\,;\quad x\,=\,p^2\,.\nonumber
\end{eqnarray}
where $F_{0G}$ means an inhomogeneous part of the equation, which in
Figure~\ref{fig:compenG} is denoted by the striped triangle. We discuss
the inhomogeneous part later on.

Assuming $F_{0G}\,=\,Const$, by successive differentiations of equation~(\ref{EQ}) we obtain a linear differential equation for F(x),
in which new variable $z$ is introduced
\begin{eqnarray}
& &\biggl[\biggl(z\frac{d}{dz} + \frac{3}{5}\biggr)\biggl(z\frac{d}{dz} + \frac{2}{5}\biggr)\biggl(z\frac{d}{dz} + \frac{1}{5}\biggr)\biggl(z\frac{d}{dz}\biggr)\biggl(z\frac{d}{dz} - \frac{1}{5}\biggr)\biggl(z\frac{d}{dz} - \frac{2}{5}\biggr)\times \nonumber\\
& &\biggl(z\frac{d}{dz} - \frac{3}{5}\biggr)\biggl(z\frac{d}{dz} - \frac{4}{5}\biggr) +z \biggl(z\frac{d}{dz} + \frac{14}{15}\biggr)\biggr] F(z)\,=\,0\,;\quad z\,=\,\frac{81\,G^2\,x^5}{15625\,\pi^2}\,.\label{z}
\label{EQD}
\end{eqnarray}

The differential equation is equivalent to equation~(\ref{EQ}) with boundary conditions. We have the following solution with an account of normalization
condition $F(0) = 1$
\begin{equation}
F(z) = \frac{6\,\Gamma\Bigl(\frac{1}{15}\Bigr)}
{125\,\Gamma\Bigl(\frac{4}{5}\Bigr)}\, G^{50}_{18}\Bigl( z |^{1/15}_{0,\, 1/5,\, 2/5,\, 3/5,\, 4/5,\, -3/5\, -2/5,\, -1/5}\Bigr);\label{solution}
\end{equation}
where
$
G^{m n}_{p q}\Bigl(\,z\,|^{a_1,\,...,\,a_p}_{b_1,\,...,\,b_q}\Bigr)\,
$
is a Meijer function~\cite{BE}.

\begin{figure}
\includegraphics[scale=1.2,width=15cm]{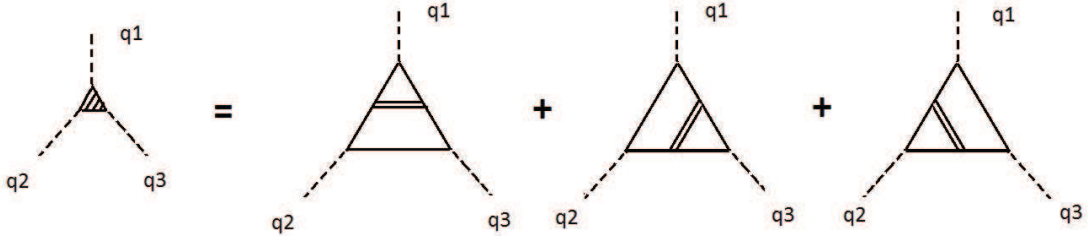}
\caption{Diagrams, describing the first approximation for the Standard Model contribution to three-graviton vertex~(\ref{GGG}). Simple lines
correspond to matter fermions (quarks {\it etc.}, double lines
correspond to weak bosons $W$.}
\label{fig:neutrino}
\end{figure}

On the other hand, assuming $F_{0G}\,=\,0$, we may calculate $F(0)$ from equation~(\ref{EQ}), that gives
\begin{equation}
F(0)\, = \, \frac{18}{5}\,.\label{eq:F008}
\end{equation}
However, form-factor $F(z)$ has to be unity at zero.
So there is evidently additional contribution to $F(0)$, that is
\begin{equation}
F_{0G}\,\ne\,0\,.\label{eq:Fne}
\end{equation}
This contribution might be given by diagrams including matter fields, for example, by those
being presented in Figure~\ref{fig:neutrino}.

First of all we would draw attention to presence of $W$ exchange in Figure~\ref{fig:neutrino}.

The
contribution is provided by the the T-odd and P-odd part of fundamental fermions interaction with weak bosons $W^\pm$.
The interaction of $W$ with quarks and leptons contains $\gamma_5$ matrix and so
corresponding traces inevitably contains antisymmetric tensor
$\epsilon_{\alpha \beta
\gamma \delta}$, which is present in interaction~(\ref{GGG}).
The vertex of a graviton interaction with a
spinor field, is the following
\begin{equation}
V(\mu,\nu,p_1,p_2) = \kappa (\gamma_\mu (p_1+p_2)_\nu+
\gamma_\nu (p_1+p_2)_\mu);\nonumber
\end{equation}
where
$p_1$ is the momentum of the incoming spinor,
$p_2$ is the
same of the outgoing one.
$\kappa$ is connected with the Planck mass
\begin{equation}
\kappa\,=\,\frac{1}{M_{Pl}}\,;\label{MP}
\end{equation}
The $T$-odd part of the quarks weak interaction is the following
\begin{equation}
-\frac{\imath\,g\,A\,\lambda^3\,\overline{\eta}}{\sqrt{2}}\Bigl(
\overline{u}\gamma_\mu(1+\gamma_5)b + \overline{t}\gamma_\mu(1+\gamma_5)d
\Bigr)\,W^\mu+\,h.c.\,;\label{Tint}
\end{equation}
where~\cite{PDG}
\begin{equation}
\lambda = 0.22537\pm0.00061;\;\overline{\eta} = 0.353\pm0.013;\;A = 0.814^{+0.023}_{-0.024}.\label{lea}
\end{equation}

We readily estimate, that diagrams in Figure~\ref{fig:neutrino} with quark loops give the following contribution to
the inhomogeneous part of the equation
\begin{equation}
F_{0G}\,=\,-\,\frac{3\,g^2 \kappa^3\,A\,\lambda^3 \overline{\eta}}{64\,\pi^4\, G\,M_W^2}\,\ln\,\frac{M_W\,m_b}{m_u\,m_d};\label{eq:kappaG8}
\end{equation}
where $g \simeq 0.65$
is the electro-weak gauge constant. As for lepton loops, we have no full information yet on corresponding mixing parameters, and thus we use for estimates only~(\ref{eq:kappaG8}), the more so, as a quark loop has additional color factor 3.
From the main equation~(\ref{EQ})
we have the following condition
\begin{equation}
F(0)\,+\,F_{0G}\,= 1\,.\label{G}
\end{equation}
Expression~(\ref{eq:kappaG8}) has to be equal to
\begin{equation}
F_{0G}\,= 1\,-\,F(0)\,=\,-3.003\,.\label{eq:G08}
\end{equation}

Then with previous relations ~(\ref{EQ}, \ref{eq:G08}) we obtain the following estimate
for the coupling constant of the effective interaction~(\ref{GGG})
$G$.
In doing this we have to bear in mind, that integral equation~(\ref{EQ}) is divided by coupling constant $G$ due to the
overall procedure for searches for non-trivial solutions of compensation equations. Thus we
have
\begin{equation}
G\,\simeq\,\frac{g^2 \kappa^3\,A\,\lambda^3 \overline{\eta}}{64\,\pi^4\,M_W^2}\,\,\ln\,\frac{M_W\,m_b}{m_u\,m_d}
\,\simeq\,3.2\cdot 10^{-67}\,GeV^{-5}\,.\label{eq:Gestim}
\end{equation}

As we have already mentioned, for the moment we can not substitute reliable
values for the average neutrino mass $m_\nu$ and mixing parameters in
analogous to~(\ref{eq:Gestim}) lepton expression. We may only safely assert,
that  $m_\nu$ is not zero due to existence of the effect of neutrino
oscillations. In any case it may not be more than $3\, eV$.

In view of this we
have taken for the estimate just quarks, as particles giving contribution to coupling constant $G$, the more so, as in the quark loops we have
the additional color factor 3. It is evident, that massless
particles, namely photons and gluons, do not give contribution due to
parity conservation of their interactions.
To obtain more definite connection between two parameters $G$ and
$\kappa$ one needs perform difficult calculations, which will be done elsewhere. However our estimate~(\ref{eq:Gestim}) allows us to consider
qualitatively effects of the interaction~(\ref{GGG}).

With physical mass of $W$ and bearing in mind relation~(\ref{MP}),
where Planck mass~$M_{Pl} \simeq 1.22\cdot 10^{19}\,GeV$ is very large, we
understand, that
possible value~(\ref{eq:Gestim}) is essentially larger, than seemingly
natural value $G_{Pl}\,\sim\, \kappa^5$, which one can estimate under premise, that only
gravitational effects can define the quantity under the study.

The interaction~(\ref{GGG}) due to a presence of the antisymmetric tensor $\epsilon_{\alpha \beta \gamma \delta}$ gives no contribution to spherically symmetric problems of gravitation (Schwartzschield solution, Friedmann
solution {\it etc.}).
However it could manifest itself in problems without
spherical symmetry in a rotating system ({\it e.g.} spiral galaxy). The considerable enhancement of possible value~$G$ in comparison
to natural value~$G_{Pl}$ by the following factor
\begin{equation}
\frac{G}{G_{Pl}}\,=\,\frac{g^2\,A\,\lambda^3\,\overline{\eta}\,M_{Pl}^2 }{64\,\pi^4\,M_W^2}\,\ln\,\frac{M_W\,m_b}{m_u\,m_d}\simeq\,8.7\cdot 10^{28}\,;\label{eq:GnaturalR}
\end{equation}
is quite remarkable and presumably may lead to observable effects.
Let us remind, that effective interaction~(\ref{GGG}) is
 $P$ and $T$ non-invariant.

\section{A model for a running gravity coupling}

We have considered above wouldbe properties of the quantum gravity.
The theory itself contains dimensional coupling constant
\begin{equation}
\beta \,=\,\frac{\kappa^2}{4\,\pi}\,.\label{beta}
\end{equation}
In a conventional quantum field theory such quantity corresponds to a
running coupling, for example, $\alpha_s(Q^2)$ in QCD. Thus one should
expect, that quantity~(\ref{beta}) is also a running one. However,
the quantum gravity theory is non-renormalizable. It means
that we have no regular method
to obtain an expression for the running coupling.

So an application of a  non-perturbative approach is necessary. As a matter of fact,
in gauge theories of the Standard Model contributions of the non-perturbative nature may be present also. We may refer just to the strong coupling $\alpha_s(Q^2)$, in which the well known non-physical Landau singularity appears in perturbative calculations. It is a general belief,
that non-perturbative contributions somehow eliminate this singularity.

In particular, in work~\cite{AZ13} we have shown, that the singularity is excluded due to the spontaneous generation of effective three-gluon
interaction being analogous to~(\ref{FFF}). In any case, a consideration of
possible properties of running gravitational coupling~(\ref{beta}) deserves an attention. In the present section we consider a model, which illustrates the
problem.

It is very important to have
some hints on the scale of the possible non-perturbative effects in
our problem.
Here the example, which is considered in the previous Section may be
instructive.

First of all, let us consider relation~(\ref{eq:GnaturalR}), which leads to
the following estimate of characteristic length $l_{eff}$ for effective
interaction~(\ref{GGG})
\begin{equation}
l_{eff} \simeq l_{Pl} \biggl(\frac{G}{G_{Pl}}\biggr)^{\frac{1}{5}}
\simeq 6.1 \cdot 10^5\cdot l_{Pl} \simeq 10^{-27}\,cm;\label{Leff}
\end{equation}
that is we have an essential enhancement of the characteristic length in
comparison to the Planck one.

However effective interaction~(\ref{GGG}) is not the only possible one. Maybe
there are also other effective interactions, which have even larger characteristic lengths.
For example, we see, that in expression~(\ref{eq:Gestim})
both maximal mass parameter $M_{Pl}\,(\kappa=1/M_{Pl})$ and masses of fundamental particles are present. Neutrino mass $m_\nu$ is the minimal one known. So we estimate a maximal possible scale of the length dimension for  non-perturbative effects as follows
\begin{equation}
l_0\,=\,\frac{M_{Pl}}
{m_\nu^2}\,=
\,2.711\cdot 10^{20} m\,=\,8.78\, kpc;\label{estimate}
\end{equation}
where we have used for the neutrino mass its upper bound $m(\nu_e) \leq 3\,eV$~\cite{PDG}. We see, that this estimate gives a $kpc$ scale, which is appropriate to a size of a galaxy. Is it possible to use such scale for
application to real astrophysical problems? In what follows we
consider a hypothetical example of a gravity coupling behavior, which uses
estimate~(\ref{estimate}).

Let us assume, that running Newton gravity coupling $G_N$, which is proportional to coupling $\beta$~(\ref{beta}) depends on a distance in the
following way
\begin{equation}
G_N(r) = G_{N0}\,F(r);\quad
F(r)\,= \,\frac{1}{3}\,\,G^{\,3\,0}_{1\,4}\biggl(\frac{r}{r_0}|^{1}_{\,0,\,
2,\,4,\,-2}
\biggr)\,+\,
7\,G^{2\,1}_{1\,4}\biggl(\frac{r}{r_0}|^{1}_{\,2,\,4,\,0,\,-2}\biggr)\,;
\label{runG}
\end{equation}
where $r_0$ is of the order the of magnitude of $l_0$~(\ref{estimate}) and
$G_{N0}$ is just the well-known Newton constant. We use Meijer functions
in the model, because in similar
problems we encounter these functions. The
Fourier transform of a Meijer function is again a Meijer function~\cite{PBM}.
\begin{figure}
\includegraphics[scale=0.6,width=12cm]{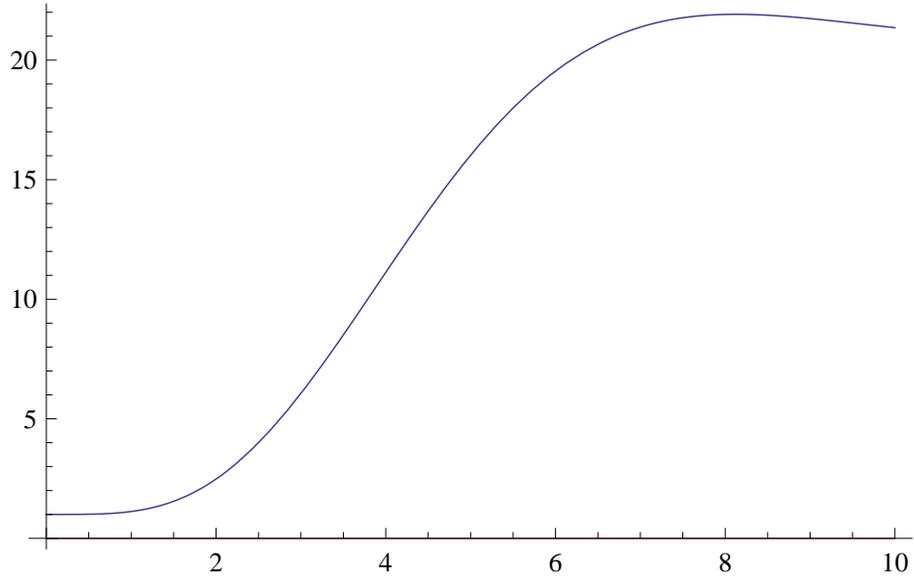}
\caption{Behavior of
$F(x^2),\, x = \sqrt{r/r_0}$.}
\label{fig:F}
\end{figure}
The gravity is not a logarithmic theory, as {\it e.g.}  QCD, but
a power-mode theory.
We choose the coefficients in~(\ref{runG})
so that for $r \to 0\;\,G_N \to G_{N0}$, and for $r \to \infty\;\,
G_N \to 21\,G_{N0}$.
The last asymptotic
corresponds to the accelerated expansion of the Universe, which usually is prescribed to a dark energy.
\begin{figure}
\includegraphics[scale=0.6,width=12cm]{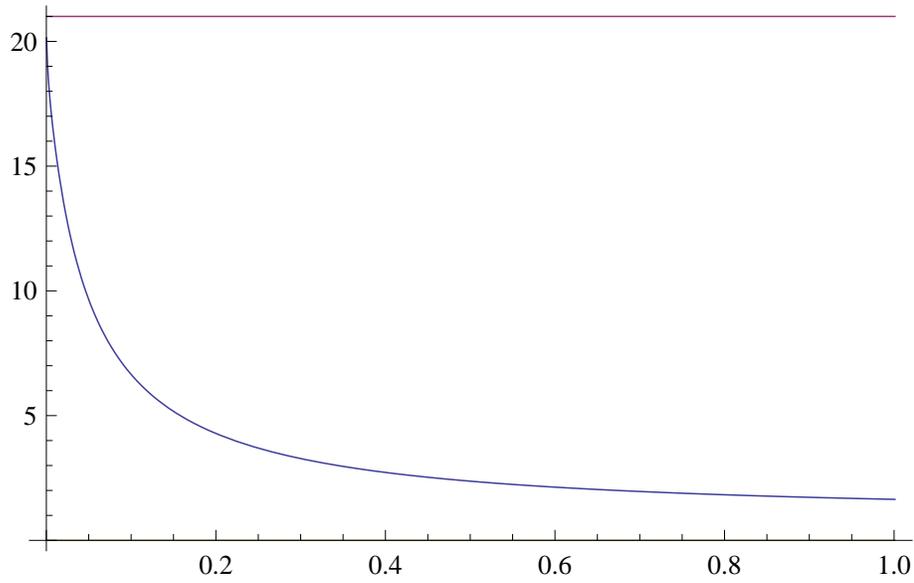}
\caption{The behavior of
$G_N(16 \,k^2 r_0^2)/G_{N0}$. The upper line corresponds to the absence of the
momentum dependence: $G_N = G_N(0) = 21\,G_{N0}$.}
\label{fig:runGk}
\end{figure}
The Fourier transform, corresponding to expression~(\ref{runG}), in a dependence on variable
$y\,=\,16\, k^2 r_0^2$, where $k$ is a momentum, is the following
~\cite{PBM}
\begin{equation}
\bar F(k^2)\,=\,\frac{4}{3\,\pi}\Bigl(G^{\,1\,4}_{6\,2}\biggl
(\,y\, |^{\,1,\,-\frac{1}{2},\,-1,\,-\frac{3}{2},\,\frac{3}{2},\,2}_{\,\frac{1}{2},\,0}
\biggr)\,+\,21\,G^{\,2\,3}_{6\,2}\biggl
(\,y|^{\,-\frac{1}{2},\,-1,\,-\frac{3}{2},\,1,\,\frac{3}{2},\,\,2}_{\,\frac{1}{2},
\,0}\biggr)\Bigr)\,.\label{fourier}
\end{equation}
Function~(\ref{fourier}) is shown in Figure~\ref{fig:runGk}. The momentum dependence resembles the situation with the asymptotic freedom in QCD. The main difference consists in a dependence at
the infinity. In QCD the strong coupling decreases
\begin{equation}
\alpha_s(k^2)_{k^2 \to \infty} \simeq \frac{4\,\pi}{\beta_0\,\ln \frac{k^2}{\Lambda^2}}\,;\label{QCD}
\end{equation}
while in the case under
the consideration the gravity coupling tends to the Newton constant
\begin{equation}
\lim\, G_N(k^2)_{k^2 \to \infty} = G_{N0}.\label{Ginfty}
\end{equation}

Now we apply representation~(\ref{runG}) to rotation curves of galaxies.
Let us choose $r_0 = 5\, kpc$.
We take for a rotation curve of a flat disk galaxy the following
expression
\begin{eqnarray}
& &V(r) =V_0 \sqrt{2 M_G (I_0(y) K_0(y)-I_1(y) K_1(y)) F(r)};\nonumber\\
& &y\,=\,\frac{r}{R_G}\,;\quad V_0\,=\,207.4\,\frac{km}{s}\,;\label{VV}
\end{eqnarray}
where $M_G$ is a galaxy mass in $10^{10}\,M(Sun)$, $R_G$ is its
radius in $kpc$ and r is a
distance in a rotation curve in $kpc$ as well.

Then we substitute a galaxy
mass and its radius to obtain a corresponding rotation curve. We have taken
five galaxies for examples. Black spots in figures denote observation
data, being taken from~\cite{sanders}.

\begin{figure}
\begin{center}
\includegraphics[scale=0.6,width=11cm]{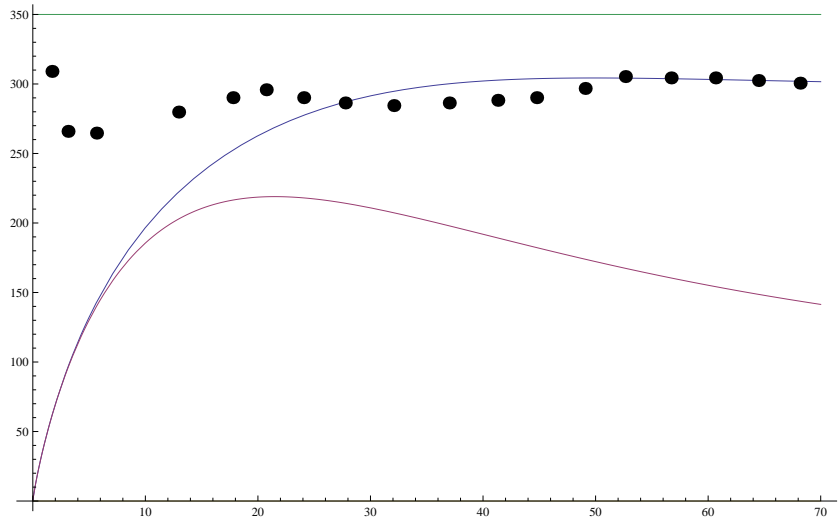}
\caption{Rotation curve for Galaxy UGS 2885.}
\label{fig:NDM30}
\end{center}
\end{figure}

\begin{figure}
\begin{center}
\includegraphics[scale=0.6,width=11cm]{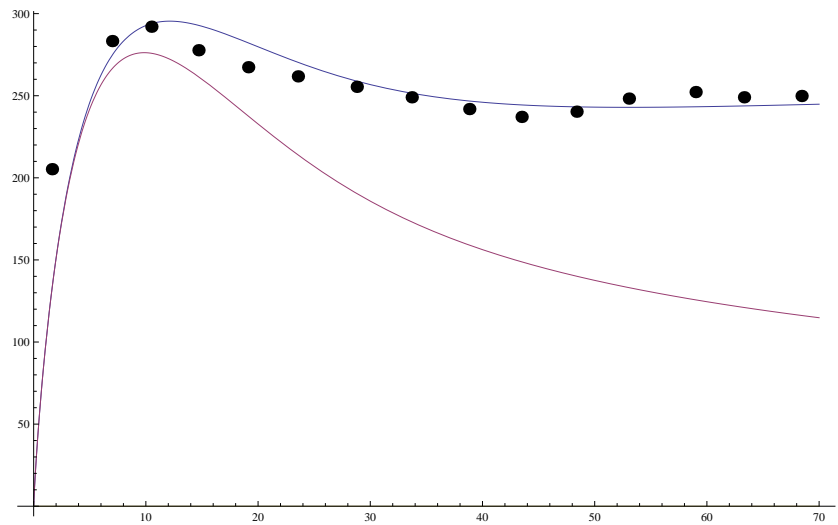}
\caption{Rotation curve for Galaxy NGC 6674.}
\label{fig:NDM21}
\end{center}
\end{figure}

\begin{figure}
\begin{center}
\includegraphics[scale=0.6,width=11cm]{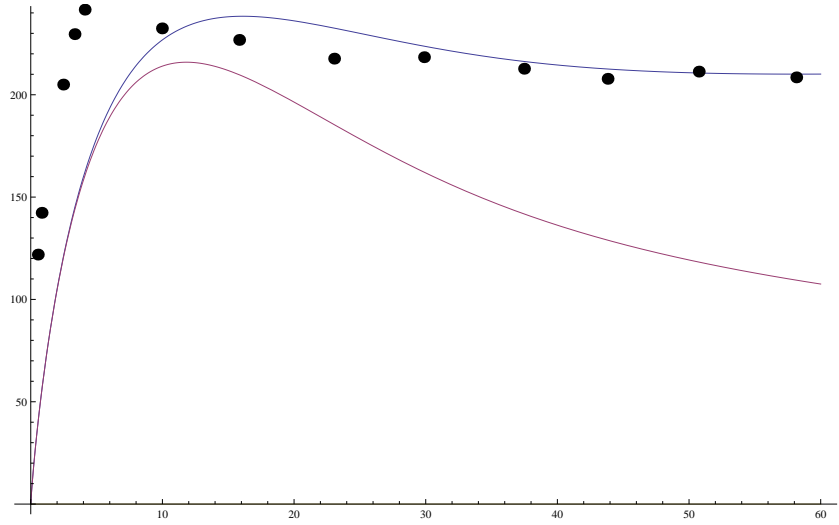}
\caption{Rotation curve for Galaxy NGC 801.}
\label{fig:NDM16}
\end{center}
\end{figure}

\begin{figure}
\begin{center}
\includegraphics[scale=0.6,width=11cm]{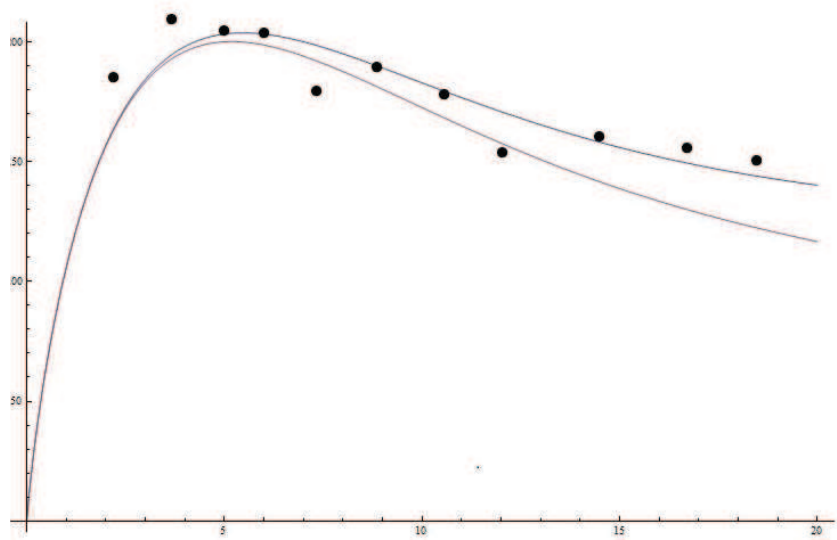}
\caption{Rotation curve for Galaxy NGC 3521.}
\label{fig:NDM9}
\end{center}
\end{figure}

\begin{figure}
\begin{center}
\includegraphics[scale=0.9,width=11cm]{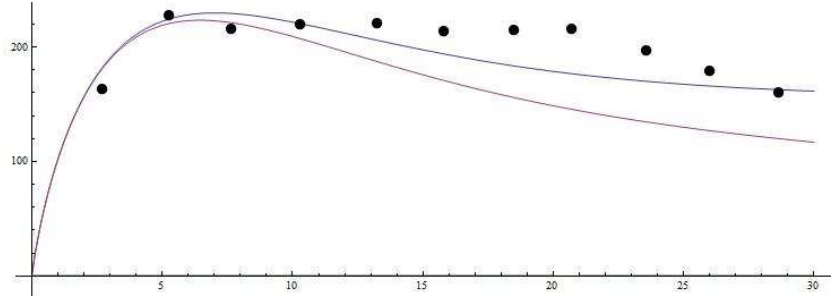}
\caption{Rotation curve for Galaxy NGC 2683.}
\label{fig:NDM58}
\end{center}
\end{figure}

Upper curves in Figures correspond to
representation~(\ref{VV}). All figures show how velocities in $km/s$ depend on distances in $kpc$.
Lower curves in Figures correspond to $F(r)=1$, {\it i.e.} without
additional interaction. The examples show, that representation~(\ref{runG})
essentially improves the agreement with observational data.\\
\\
Galaxy UGS 2885, $M_G = 28.8,\,R_G = 10$, Figure~\ref{fig:NDM30}.\\
Galaxy NGC 6674, $M_G = 21.0,\,R_G = 4.58$, Figure~\ref{fig:NDM21}.\\
Galaxy NGC 801, $M_G = 15.4,\,R_G = 5.5$, Figure~\ref{fig:NDM16}.\\
Galaxy NGC 3521, $M_G = 9.0,\,R_G = 3.0$, Figure~\ref{fig:NDM9}.\\
Galaxy NGC 2683, $M_G = 5.8,\,R_G = 2.4$, Figure~\ref{fig:NDM58}.


\newpage
\section{Conclusion}

The present work deals with a possibility of a presence of effective interactions in
the gravity theory, the fundamental General Relativity. What are benefits of such possibility?\\
1. A wouldbe appearance of the P-odd and T-odd effective
interaction of gravitons with a new scale.
\\
2. The toy model for dark effects, which with an assumption on a
special scale can pretend to a description of the real situation in
the astrophysics.

As for the last item, it is for the moment a hypothesis. It just show a possibility how things might be, in case
searches for a dark matter and a proper understanding of the dark
energy would fail. Note, that numerous proposals for a physical content of
the dark substances also have a hypothetical status yet.

There are three regions in this model, in which properties of the gravitation
turn to be quite different. In the first region $r \ll 10\,kpc$ the gravity
coupling with the high accuracy equals the Newton one
 $G_{N0}$. In the region with
$r \simeq 10\,kpc$ this coupling increases, that leads to effects in the rotation curves, which usually are prescribed to a dark matter. Finally, in the region with $r \gg 10\,kpc\;G_N$ is 21 times  more, than the Newton coupling $G_{N0}$, that might serve as an explanation of the accelerated expansion of the Universe, which is nowadays usually prescribed to the dark energy.  Thus, in case of such dependence of the gravity on a distance being realized, the presumptive effects of the dark matter and the dark energy are simultaneously described. Of course the present results
are qualitative, a further specification is necessary for a more accurate
comparison with data.

The purpose of the work
is just to demonstrate a wouldbe tool to deal with effects in
the gravity physics, which relies on close similarity of the gravity and
vector non-abelian fields.

\section{Acknowledgments}

The work is supported in part by the Russian Ministry of Education and Science
under grant NSh-3042.2014.2.

\end{document}